\documentclass[showpacs,preprintnumbers,amsmath,amssymb]{revtex4}

\usepackage{graphicx}
\usepackage{dcolumn}
\usepackage{bm}

\begin{document}
\def\bbox#1{\hbox{\boldmath${#1}$}}
\def\blambda{{\hbox{\boldmath $\lambda$}}}
\def\eeta{{\hbox{\boldmath $\eta$}}}
\def\bxi{{\hbox{\boldmath $\xi$}}}

\title{ Molecular States of Heavy Quark Mesons}

\author{Cheuk-Yin Wong}

\affiliation{
Physics Division, Oak Ridge National Laboratory, Oak Ridge, TN
37831, U.S.A.
}

\date{\today}

\begin{abstract}

We explore molecular states of two open heavy-quark mesons $(Q\bar
q)$-$(q\bar Q)$ in a quark-based model in terms of a four-body
non-relativistic Hamiltonian with pairwise effective interactions.
Molecular states are found in the combinations of $\{ D, D^*, B,
B^*\}$ with $\{\bar D, \bar D^*, \bar B, \bar B^*\}$, including a
weakly-bound $D\bar D^*$ state near the threshold which may be
qualitatively identified as the 3872 state observed recently by the
Belle Collaboration.

\end{abstract}

\pacs{12.39.Jh,12.39.Pn} 

\maketitle

\section{Introduction}

The narrow 3872 MeV state recently discovered by the Belle
Collaboration \cite{Bel03}, and subsequently confirmed by the CDFII
Collaboration \cite{Bau03}, has generated a great deal of interest.
The state contains a $c\bar c$ pair and was observed in the exclusive
decay of $B^{\pm} \to K^+\pi^+\pi^- J/\psi$.  Possible explanations of
the state include a conventional charmonium state, hybrid state, and
multi-quark molecular state  
\cite{Tor03,Clo03,Pak03,Vol03,Bib03,Eic03,Cha03,Bra03,Swa03,Bar03a}.
Direct experimental search using $e^+ e^-$ annihilation by the BES
Collaboration in the BEPC indicates that this state is unlikely to be
a vector $1^{--}$ state \cite{Yua03}.  The 3872 state lies about 60
MeV higher than the weighted average of the conventional $\Psi(1D)$
charmonium state predicted by the Cornell potential \cite{Eic80} and
the Buchm\"uller-Tye potential
\cite{Buc81}.  The identification of the 3872 state as $\Psi(^3D_{2})$
would imply too large a splitting between $\Psi(^3D_{2})$ and
$\Psi(^3D_{1})$, as the latter state has been identified with the 3770
state.  Furthermore, the 3872 state is seen in the $\pi^+\pi^-$
transition to $J/\psi$ and only an upper bound on E1 transition to the
$\chi(^3P_J)$ state can be determined experimentally, whereas the
$\Psi(^3D_{2})$ state would have yielded a large partial E1 transition
width, about 5 times greater than the width for the $\pi^+\pi^-$
transition \cite{Eic02}.  On the other hand, the 3872 state lies very
near the $D^{*0}\bar D^0$ threshold. The proximity of this state near
the threshold and the difficulties with the $\Psi(^3D_{2})$
description led the Belle Collaboration to suggest that the observed
state may be a multi-quark `molecular state' studied in previous
theoretical investigations
\cite{Tor03,Clo03,Pak03,Vol03,Bib03,Eic03,Cha03,Bra03,Swa03,Ruj77,Jaf76,Dol74,Liu80,Won80,Wei82,Tor92,Man92,Eri93,Wei90,Bar94,Bar03,Doo92,Lip77,Ade82,Hel85,Bri98,Gre97,Mih97,Mic99,Bar99,Ric02}.

We shall be interested in molecular states of the type $(Q\bar
q)$-$(q\bar Q)$ where $Q$ is a charm or bottom quark and $q$ is a $u$
or $d$ quark.  We shall often use the simplified nomenclature of ``$M$
and $\bar M^*$'' or $M\bar M^*$ to denote mesons
$M$ and $ \bar M^*$ or their charge conjugate $M^*$ and $\bar M$.
Molecular states formed by $D$ and $\bar D^*$ were predicted many
years ago by T\" ornqvist \cite{Tor92} who suggested that open
heavy-quark pairs form deuteron-like meson-meson states, `deusons',
because of the strong pion exchange interaction.  In the heavy meson
sector, T\" ornqvist estimated that one-pion exchange alone is strong
enough to form deuteron-like composites of $B\bar B^*$ and $B^* \bar
B^*$ bound by about 50 MeV, and composites of $D\bar D^*$ and $D^*
\bar D^*$ bound by pion exchange alone are expected near the
threshold.  Manohar and Weise have also studied another type of
multi-quark systems, the $(Q Q
\bar q \bar q)$ hadronic states, in the limit where the quark mass
$m_Q$ of $Q$ goes to infinity.  They noted that there is a long-range
binding due to the one-pion exchange between ground state $Q \bar q$
mesons.  They suggested that for two open bottom mesons, this
long-range interaction may be sufficient to produce a weakly-bound
two-meson state \cite{Man92}.  Ericson and Karl discussed the physics
of pion exchange and the role of tensor forces in forming such
hadronic molecules \cite{Eri93}.

The pion-exchange model presents a reasonable description of the
long-range attraction between $D$ and $\bar D^*$.  However, as the
state energy depends on the strength of the potential at short
distances, the results of the pion-exchange model depends on how the
long-range potential is regularized at short distances. The results
may also be affected when exchanges of more mesons are included.

To describe the short-distance behavior, it is desirable to study
multi-quark molecular states in a quark-based model.  The
phenomenological interaction between a quark and an antiquark in a
single meson is reasonably well known
\cite{Eic80,Buc81,God85,Bar92,Swa92,Won02,Bar03b,Won00}.
However, there are considerable differences and uncertainties in the
description of the interaction between constituents in a baryon, as it
has been given in terms of one-gluon exchange (OGE) interactions
involving color operators of the type $\bbox{\lambda}(i)
\cdot\bbox{\lambda}(j)$
\cite{Wei90,Bar94,Bar03,Doo92,Lip77,Ade82,Hel85,Bri98,Gre97,Mih97,Mic99,Bar99,Ric02},
or alternatively in terms of a one-boson meson exchange (OBE)
interactions involving flavor operators of the type $\bbox{\tau}(i)
\cdot\bbox{\tau}(j)$ \cite{Glo96,Fuj96,Glo98,Liu99,Sta03}.  
Here, $\bbox{\lambda}(i)$ is the generator of the SU(3)$_{\rm color}$
group for particle $i$ and $\tau(i)$ is the generator of the
SU(2)$_{\rm flavor}$ group.  The interquark interactions in the OGE
and OBE models have very different color structures.

For the interaction between constituents of different mesons, Lipkin
and Greenberg pointed out pathological problems involving the use of
confining one-gluon exchange interactions in a quark-based model
because they do not respect local color gauge invariance
\cite{Lip82,Gre81}.  The underlying difficulty arises because the
quark-based one-gluon exchange model represents a truncation of the
basis states with the neglect of the dynamics of explicit gluon
degrees of freedom. In spite of these defects, the quark-based
one-gluon exchange model with the interchange of quarks has been used
successfully to study the interaction of mesons at short distances
\cite{Wei90,Bar92,Swa92,Won02,Bar03b,Won00} when the dynamics of the
gluon degrees of freedom can be neglected.

In a recent quark-based model study of molecular states using the
one-gluon exchange model with constituent interchanges, Swanson found
that the one-gluon exchange interaction couples $D \bar D^*$ with
$\omega (J/\psi)$, but the interaction is not attractive enough to
lead to a bound molecular state \cite{Swa03}.  Swanson then studied
molecular states with the meson-based one-pion exchange model (as in
T\"ornqvist
\cite{Tor92}) and added the quark-based one-gluon exchange
interaction with constituent-interchanges.  The probability of the
mixing of $\omega (J/\psi)$ with $D \bar D^*$ depends on the one-pion
exchange potential cut-off parameter $\Lambda$, ranging from zero for
the cut-off value of $\Lambda$ used by T\" ornqvist without the
one-gluon exchange interaction \cite{Tor92} to a maximum mixing of
17\% at large values of $\Lambda$.  In such a treatment in which
the interaction between quarks is superimposed on the one-pion
exchange interaction between mesons, the position of the molecular
state relative to the $D\bar D^*$ state and the degree of $\omega
(J/\psi)$ mixing depends on how the long-range pion exchange potential
is regularized at short distances and the result may also be affected
when exchanges of more mesons are included.

We search for a description of the heavy-quark molecular states within
a completely quark-based model.  We note that the physical conditions
appropriate for the molecular state of our interest have important
effects on the interaction between the constituents in different
mesons.  The molecular state at the center of our attention has a
binding energy of a few MeV.  For such a weakly-bound molecular state,
the average separation $R$ between the heavy mesons is considerably
greater than the average radius of heavy quark mesons $a$ (See Section
III for more detail), as was already recognized by Close and Page,
Voloshin, and Braaten and Kusunoki \cite{Clo03,Vol03,Bra03}.  At these
large distances with $R >> a$, the probability for the recombination
of a quark of one meson with the antiquark of the other meson after a
gluon exchange is highly suppressed as their quark-antiquark
separation $R$ is much greater than their average natural meson radius
$a$.  When $R >> a$, far more likely after the exchange of a single
gluon is the occurrence of the exchange of an additional gluon between
quarks.  This two-gluon emission leads to the color van der Waals
interaction proportional to the inverse-power of $R$, as shown by
Applequist $et~al.$, Peskin $et~al.$, Bhanot $et~al.$, Lipkin,
Greenberg, and many other authors
\cite{App77,Pes79,Bha79a,Bha79b,Lip82,Gre81,others}.  The color van
der Waals interaction can be equivalently represented in terms of
effective charges of quarks and antiquarks in a QED-type
interaction. There is however no experimental evidence for the color
van der Waals interaction of the inverse-power type between separated
hadrons. As pointed out by Lipkin and Greenberg
\cite{Lip82,Gre81}, it is necessary to include additional dynamics of
gluons in such a quark-based model.  Accordingly, due to the breaking
of the gluonic string at large distance, the color van der Waals
interaction should be screened by introducing a screening mass $\mu$,
which will modify the inverse-power color van der Waals interaction at
large distances. Appropriate for the peculiar physical conditions of
weakly bound molecular states, the concepts of effective charges and
the screening mass are incorporated into the effective interquark
potential which is subsequently used to explore meson-meson molecular
states.

Nucleus-nucleus molecular states have been observed previously in the
collision of light nuclei near the Coulomb barrier
\cite{Alm60,Kon79,Sat84}.  Effective nucleon-nucleon interactions have
been successfully applied to obtain the potential between two nuclei
to study their reactions \cite{Sat84}.  In a similar way, effective
interquark potentials can be used to obtain the potential between
mesons.  Just as with the nucleus-nucleus potential, the meson-meson
potential can be evaluated as a sum of a direct potential and a
polarization potential.  The direct potential arises from the
interaction of a constituent of one meson with a constituent of
another meson, and the polarization potential arises when one meson
polarizes the other meson in its vicinity.  Using Gaussian wave
functions for the mesons, the direct potential and the polarization
potential can be obtained analytically.  The knowledge of the
potential between the mesons then allows one to determine the
eigenstate of the four-quark system.

Using the effective interaction, we find many weakly-bound molecular
states in $(Q \bar q)$-$(q \bar Q)$ systems with charm and bottom
quarks, including a weakly-bound $D^+  D^{*-}$ state near the threshold
which may be qualitatively identified as the 3872 state.  For the
bottom meson $(Q \bar q)$-$(q \bar Q)$ pairs, there are weakly-bound
2S states near the threshold, in addition to 1S states with binding
energies of about 150 MeV.

Our exploratory study using effective charges and effective
interactions in a non-relativistic quark model differs from those of
the pion-exchange models and the multi-quark models with explicit
single-gluon or single-boson exchanges. The results of molecular
states obtained from different models can naturally be different.  A
careful comparison of these different results with experimental data
will be useful to determine the importance of various mechanisms that
are present in generating the molecular states.

This paper is organized as follows.  In Section II, we start with the
four-quark Hamiltonian and partition the Hamiltonian into an
unperturbed part of two mesons and a residue interaction.  The
eigenvalue equation for the four-quark system is reduced to a Schr\"
odinger equation for the relative motion of the two mesons in a
meson-meson potential.  In Section III, we present justifications
for the introduction of the effective charges and the screening mass
in the effective interquark interaction.  In Section IV, we present
our method for the evaluation of the molecular potential.  Section V
presents the results of the meson-meson potential and eigenvalues for
four-quark states.  We find molecular states in the combinations of
$\{ D, D^*, B, B^*\}$ with $\{\bar D, \bar D^*, \bar B, \bar B^*\}$.
In Section VI, we discuss the origin of the short-distance attraction
in $(Q\bar q)$-$(q\bar Q)$ systems and contrast them with the
short-distance repulsion in $(Q\bar q)$-$(Q\bar q)$ systems. In
Section VII we present our conclusions and discussions.

\section{Hamiltonian for the four-quark system}

To study the $(Q\bar q)$-$(q\bar Q)$ system, we label constituents $Q$,
$\bar q$, $q$, and $\bar Q$ as particles 1, 2, 3, and 4 respectively
and describe the four-quark system with a non-relativistic Hamiltonian
\begin{eqnarray}
\label{eq1}
H=\sum_{j=1}^4 \frac{ \bbox{p}_{j}^2} { 2 m_j} + \sum_{j=1}^4 \sum_ {k>j}^4
V_{jk} +\sum_{j=1}^4 m_j,
\end{eqnarray}
in which particle $j$ has a momentum $\bbox{p}_j$ and a rest mass
$m_j$.  The pairwise interaction $V_{jk}(\bbox{r}_{jk})$ between
particle $j$ and particle $k$ depends on the relative coordinate between
them,
\begin{eqnarray}
\label{rjk}
\bbox{r}_{jk}=\bbox{r}_j-\bbox{r}_k. 
\end{eqnarray}
We introduce the two-body momentum 
\begin{eqnarray}
{\bbox P}_{jk}=\bbox
{p}_j+\bbox{p}_k,
\end{eqnarray}
and the two-body internal relative momentum
\begin{eqnarray}
\label{e1}
\bbox{p}_{jk}=f_k \bbox{p}_j-f_j \bbox{p}_k,
\end{eqnarray}
where 
\begin{eqnarray}
\label{e2}
f_k=m_k/m_{jk},
\end{eqnarray}
\begin{eqnarray}
\label{e3}
m_{jk}=m_j+m_k.
\end{eqnarray}

There are many ways to partition the total Hamiltonian $H$ into an
unperturbed two-meson part and a residue interaction $V_I$.  We
partition the Hamiltonian so that in the lowest order the state in
question can be described by a state of the unperturbed Hamiltonian.
We can, for example, choose to partition $H$ into an unperturbed
Hamiltonian $h_{12}+h_{34}$ of mesons $A(12)$ and $B(34)$ according to
\begin{eqnarray}
H=\frac{ \bbox{P}_{12}^2}{2 m_{12}} + \frac{ \bbox{P}_{34}^2}{2
m_{34}} +V_I + h_{12} + h_{34} ,
\end{eqnarray}
\begin{eqnarray}
V_I=V_{14}(\bbox{r}_{14})+V_{13}(\bbox{r}_{13})
+   V_{23}(\bbox{r}_{23})+V_{24}(\bbox{r}_{24}),
\end{eqnarray}
\begin{eqnarray}
h_{jk}=\frac{ \bbox{p}_{jk}^2}{2 \mu_{jk}} +
V_{jk}(\bbox{r}_j-\bbox{r}_k) + m_{jk} {\rm
~~~~for~~}(jk)=A(12){\rm~and~}B(34),
\end{eqnarray}
where $\mu_{jk}=m_j m_k/m_{jk}$.  The eigenvalues of the Hamiltonians
$h_{12}$ and $h_{34}$ can be solved separately to obtain the bound
state wave functions and masses $M_{jk}(\nu)$ of mesons $A$ and $B$,
\begin{eqnarray}
h_{jk}|(jk)_\nu\rangle =
[ \epsilon_{jk}(\nu) + m_{jk}] |(jk)_\nu\rangle
=M_{jk}(\nu)  |(jk)_\nu\rangle.
\end{eqnarray}
The four-body Hamiltonian becomes
\begin{eqnarray}
H=\frac{ \bbox{P}_{12}^2}{2 m_{12}} + \frac{ \bbox{P}_{34}^2}{2
m_{34}} +V_I + M_{12}(\nu)+ M_{34}  (\nu') .
\end{eqnarray} 
The above non-relativistic approximation is obtained from relativistic
results by neglecting terms of order $\epsilon_{jk}/M_{jk}$ and higher
\cite{Won01}.  In order to satisfy the boundary condition at large
separations for which $V_I$ approaches zero, we need to include some
of these higher-order terms and modify $m_{12}$ and $m_{34}$ in the
above equation to $M_{12}(\nu)$ and $M_{34}(\nu')$ so that the
Hamiltonian becomes
\begin{eqnarray}
H=\frac{ \bbox{P}_{12}^2}{2 M_{12}(\nu)} 
+ \frac{ \bbox{P}_{34}^2}{2 M_{34}(\nu')}
+V_I + M_{12}(\nu)+ M_{34}  (\nu'). 
\end{eqnarray} 
The above Hamiltonian then describes properly the asymptotic behavior
at large separations of the two mesons.

To solve for the eigenstates, we use the two-body meson states
$|A_\nu(12) B_{\nu '}(34) \rangle$ as basis states.  We consider a
four-quark system in which the mesons would be in the $|A(12)_\nu
B(34)_{\nu'}
\rangle$ state if there were no residue interactions $V_I$.  When we
include the residue interactions $V_I$ as a perturbation, the
eigenfunction of $H$ becomes
\begin{eqnarray}
\Psi(\bbox{r},\bbox{r}_{12},\bbox{r}_{34})
=\psi(\bbox{r})\left \{|A_\nu B_{\nu'}\rangle
-{\sum_{\lambda,\lambda'}}' \frac{|A_{\lambda} B_{\lambda'}\rangle
\langle A_{\lambda} B_{\lambda'} | V_I | A_\nu B_{\nu'} \rangle}
{\epsilon(A_{\lambda})+\epsilon(B_{\lambda'})- 
\epsilon(A_\nu)-\epsilon(B_\nu')} \right \},
\end{eqnarray}
where 
\begin{eqnarray}
\bbox{r}= \bbox{R}_{12}-\bbox{R}_{34}, 
\end{eqnarray}
and $\bbox{R}_{jk}$ is the center-of-mass coordinate of $m_j$ and
$m_k$,
\begin{eqnarray}
\label{e4}
\bbox{R}_{jk}=f_j\bbox{r}_j+f_k \bbox{r}_k.
\end{eqnarray}
The spatial coordinates of the four-particle system and
$\bbox{r},\bbox{r}_{jk},$ and $\bbox{R}_{jk}$ are shown in Fig. 1.

\begin{figure} [h]

\includegraphics[scale=0.60]{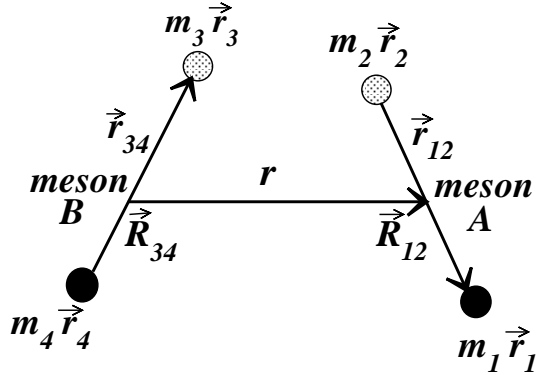}

\caption{ The spatial coordinates of the four-particle system.
}
\end{figure}

The wave function $\psi(\bbox{r})$ describes the 
relative motion between the two mesons and the symbol $\sum_{\lambda
\lambda'}'$ is to indicate that the sum is over all meson states
except $| A_{\nu} B_{\nu'}\rangle$.  The eigenvalue equation is
\begin{eqnarray}
H\Psi (\bbox{r},\bbox{r}_{12},\bbox{r}_{34})
=[M_{12}(\nu)+M_{34}(\nu')+\epsilon] 
 \Psi (\bbox{r},\bbox{r}_{12},\bbox{r}_{34}).
\end{eqnarray}
Working in the center-of-mass frame and taking the scalar product of
the above equation with $|A_{\nu} B_{\nu'}\rangle$, we obtain the Schr\"
odinger equation for relative motion of mesons $A_{\nu}(12)$ and
$B_{\nu'}(34)$,
\begin{eqnarray}
\label{sch}
\left \{\frac{\bbox{p}^2}{2 \mu}_{AB} + V(r) \right \} \psi(\bbox{r})
=\epsilon \psi(\bbox{r}),
\end{eqnarray}
where $\bbox{p}$ is the relative momentum
\begin{eqnarray}
\bbox{p}=
\frac {M_{34}(\nu')\bbox{P}_{12}-M_{12}(\nu)\bbox{P}_{34}}
      {M_{12}(\nu)+ M_{34}(\nu')},
\end{eqnarray}
and $\mu_{AB}$ is the reduced mass of the two mesons
\begin{eqnarray}
\mu_{AB}=\frac {M_{12}(\nu)M_{34}(\nu')}
          {M_{12}(\nu)+M_{34}(\nu')}.
\end{eqnarray}
The meson-meson potential $V(\bbox{r})$ is given by
\begin{eqnarray}
V(\bbox{r})=\langle A_\nu B_{\nu'} | V_I | A_\nu
B_{\nu'} \rangle -{\sum}_{\lambda,\lambda'}'\frac{ |\langle A_{\lambda}
B_{\lambda'} | V_I | A_\nu B_{\nu'} \rangle|^2}
{\epsilon(A_{\lambda})+\epsilon(B_{\lambda'})
-\epsilon(A_{\nu})-\epsilon(B_{\nu'})}.
\end{eqnarray}
We shall call the first term the direct potential 
\begin{eqnarray}
\label{dir}
V_{\rm dir}(\bbox{r})=\langle A_\nu(12) B_{\nu'}(34) |
V_I(\bbox{r},\bbox{r}_{12},\bbox{r}_{34}) 
| A_\nu(12) B_{\nu'}(34) \rangle.
\end{eqnarray}
It arises by the direct interaction of the quark matter densities of the projectile
meson with the quark matter density of the target meson mediated by
the residue interaction $V_I$.  We shall call the second term the
polarization potential
\begin{eqnarray} 
\label{polar}
V_{\rm pol}(\bbox{r})= -{\sum}_{\lambda,\lambda'}'\frac{
|\langle A_{\lambda}(12) B_{\lambda'}(34) | 
V_I(\bbox{r},\bbox{r}_{12},\bbox{r}_{34}) 
| A_\nu(12)  B_{\nu'}(34) \rangle|^2}
{\epsilon(A_{\lambda})+\epsilon(B_{\lambda'})
-\epsilon(A_{\nu})-\epsilon(B_{\nu'})}.
\end{eqnarray}
It is always attractive and it arises from the excitation of the
colliding mesons into intermediate states due to the residue
interaction $V_I$, as in the polarization of a meson in the
vicinity of another meson.  When the potential $V(\bbox{r})$ is
determined and the eigenvalue $\epsilon$ of the Schr\"odinger equation
[Eq.\ (\ref{sch})] is obtained, the mass of the four-quark system will
be given by $M=M_{12}(\nu)+M_{34}(\nu')+\epsilon$.

\section{Effective Interactions}

We shall consider the lowest energy state for which the unperturbed
mesons are in their ground states with $\nu=\nu'=0$.  The energy of
the four-quark system depends on the interquark interaction.  The
low-energy properties of isolated mesons can be described reasonably
well by a quark and an antiquark interacting with a phenomenological
confining one-gluon exchange interaction
\cite{Eic80,Buc81,God85,Bar92,Swa92,Won02,Bar03b,Won00}.  When
generalized to the multi-quark interactions involving two
quark-antiquark pairs, the one-gluon exchange interaction contains a
color operator of the type $\bbox{F}(i) \cdot\bbox{F}(j)$, where
$\bbox{F}(i)$ is $\bbox{\lambda}(i)/2$ for a quark and is
$(-\bbox{\lambda}^\dagger(i)/2)$ for an antiquark.  It was realized by
Lipkin and Greenberg that such a quark-based one-gluon exchange model
violates local color gauge invariance because of the neglect of the
gluon dynamical degrees of freedom \cite{Lip82,Gre81}.  Nevertheless,
the quark-based one-gluon exchange model has been used successfully to
study the interaction between mesons at short distances
\cite{Wei90,Bar92,Swa92,Won02,Bar03b,Won00} when the 
gluon dynamical degrees of freedom can be neglected.

\begin{figure} [h]

\vspace*{-2.1cm}
\includegraphics[scale=0.60]{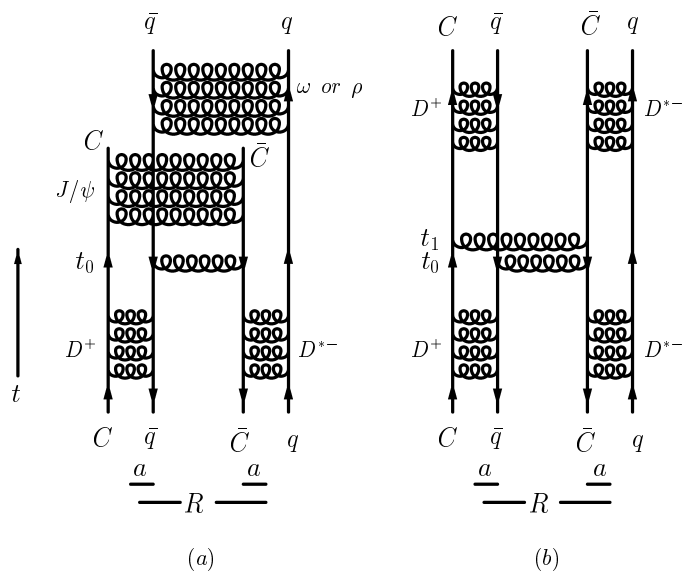}

\vspace*{-8.5cm}
\caption{ Two different ways to neutralize the color after
exchanging a gluon at time $t_0$: ($a$) Color can be neutralized by
the $C$ interacting with the $\bar C$ to form a $J/\psi$, and the $q$
interacting with the $\bar q$ to form a $\omega$ or $\rho$ . ($b$)
Color can be alternatively neutralized by exchanging another gluon at
$t_1$.  The multi-gluon exchanges in the initial and final states
represent schematically non-perturbative QCD interactions leading to
bound meson states. $R$ is the separation between mesons and $a$ is
the average meson radius.}
\end{figure}

When one applies the one-gluon exchange model to study the interaction
between mesons, the two mesons become color octet states after a gluon
is exchanged between them.  The colors of the two mesons can be
neutralized when a quark of one meson interacts with the antiquark of
the other meson to form a bound color-singlet state (Fig.\ 2$a$), as
described by the quark interchange model of Barnes and Swanson
\cite{Bar92,Swa92}.  When this quark-based model is applied to study
heavy-quark molecular states, Swanson found that the one-gluon
exchange interaction couples $D \bar D^*$ with $\omega (J/\psi)$, but
the interaction is not attractive enough to lead to a bound molecular
state \cite{Swa03}.  Swanson then studied molecular states with the
meson-based one-pion exchange model and added the quark-based
one-gluon exchange interaction with constituent-interchanges.  He
found that the probability of the mixing of $\omega (J/\psi)$ with $D
\bar D^*$ depends on the one-pion exchange potential cut-off parameter
$\Lambda$, ranging from zero for the cut-off value of $\Lambda$ used
by T\" ornqvist without the one-gluon exchange interaction
\cite{Tor92} to a maximum mixing of 17\% at large values of $\Lambda$.

To study heavy-quark molecular states within a completely quark-based
model, we note that the physical conditions associated with the
molecular state of our interest have important effects on the
interquark interaction. We consider, for example, the molecular state
of $D^+D ^{*-}$ to be a possible description of the 3872 MeV
state. It is then a weakly-bound molecular state with a binding energy
$\epsilon$ given by $\epsilon=[3.872 {\rm
GeV}-M({D+})-M({D^{*-}})]=7.4$ MeV.  The average separation $R$
between $D^+$ and $D^{*-}$ is of the order of $R\sim
\hbar/\sqrt{2 m_{\rm red} \epsilon}$ where $m_{\rm red}$ is the
reduced mass, $M({D^+})M({D^{*-}})/(M({D+})+M({D^{*-}}))$. For the
binding energy of 7.4 MeV, we can estimate the average separation $R$
between $D^+$ and $D^{*-}$ to be about 1.6 fm.  This separation $R$ is
much greater than the average radius $a$ of $D^+$ and $D^{*-}$, whose
root-mean-square radius is about 0.3 fm as given by half of the
calculated root-mean-square quark-antiquark separation listed in the
Appendix of \cite{Won02}.  Our molecular state is therefore
characterized by an average meson-meson separation $R$ much greater
than the average radius $a$ of the mesons.  This peculiar feature of
$R>>a$ has already been recognized by Close and Page, Voloshin,
and Braaten and Kusunoki \cite{Clo03,Vol03,Bra03}.

After a gluon is exchange between $D^+$ and $D^{*-}$ at $t_0$,
the $C$ and $\bar C$ are separated by a distance of the order of the
meson-meson separation $R$, as depicted in Fig.\ 2$a$.  On the other
hand, a bound $C$ and $\bar C$ will have a radius of order $a$.  When
$R >> a$, there is a mis-match between $R$ and $a$. Because of this
mis-match, the probability for the $C$ to interact with the $\bar C$
to form a bound color-singlet $C\bar C$ after the exchange of a single
gluon, as depicted in Fig. 2$a$, is highly suppressed.  The estimate
of Swanson\cite{Swa03} on the mixing of the quark-interchange
component of Fig.\ 2$a$ (using a superposition of quark-based and
meson-based models) gives a mixing probability ranging from zero
percent to a maximum of 17 percent.  For $R>>a$, this one-gluon
exchange contribution as represented by Fig 2$a$ is not the dominant
process.  Within a quark-based model, far more likely after the
exchange of a single gluon is the exchange of an additional gluon from
one meson to the other meson to neutralize the octet colors, as
depicted in Fig.\ 2$b$.  For our case of $R>>a$, it is reasonable to
consider only Fig.\ 2$b$ as the dominant contribution in the present
exploratory work.  The contribution of Fig. 2$a$ can be included in
future refinements.

The time scale for a meson to be in a color-octet state after emitting
a single gluon, which can be called the color-octet persistence time,
is of order $\hbar/(\alpha_s/a)$ \cite{Pes79}.  The time scale for
molecular motion is of order $\hbar/\epsilon$.  Therefore, the
color-octet persistence time is much shorter than the time for
molecular motion.  As a consequence, mesons undergoing molecular
motion are predominantly in their color-singlet state.  We can invoke
arguments analogous to those used previously by Applequist $et~al.$
\cite{App77} and Peskin and Bhanot $et~al.$ \cite{Pes79,Bha79a,Bha79b}
to treat the many gluons emitted from a meson to the other meson
during the short color-octet persistence period as grouping together
into clusters, and each multi-gluon cluster summing to a total
color-singlet.

Peskin evaluated the set of 10 two-gluon emission diagrams which occur
in the short color-octet persistence period \cite{Pes79}.  He showed
that in order to obtain gauge invariant results, the gluons
originating from a meson are emitted not only from the quark and the
antiquark of the meson, but also from the gluons exchanged between the
quark and the antiquark within the meson.  After the QCD two-gluon
emission diagrams from a meson have been summed, the coupling of
gluons to a static meson source is similar to the familiar coupling of
QED photons through simple QED electric dipole interactions.  The
interaction between mesons can be approximated by a dipole-dipole
interaction as in the van der Waals interaction between electric
dipoles \cite{Pes79,Bha79a,Bha79b}.  The only new feature is that the
quark and antiquark of the meson is bound into a dipole pair with
effective charges for the quark and the antiquark.  These similarities
will allow us to introduce the effective charges in a QED-type
interaction to describe the interaction between constituents in
different mesons.

We can examine the results of Peskin and Bhanot to motivate our
introduction of the effective charges.  In QCD, after the 10 two-gluon
emission diagrams have been added together, Peskin \cite{Pes79} and
Bhanot and his collaborators \cite{Bha79a,Bha79b} obtained the
gauge-invariant result that the square of the color-electric field
strength at a distance $r$ changes from the $r^{-7}$ behavior at very
large distances to a $r^{-6}$ behavior at intermediate distance, in the
same way as the Casimir-Polder effect in atomic physics \cite{Cas47}.
At a distance $r$ with $ a/\alpha_s > r > a$, in the vicinity of a
color-singlet quark-antiquark meson represented by a Wilson loop, the
square of the color-electric field strength is
\cite{Pes79,Bha79a,Bha79b}
\begin{eqnarray}
\label{ext}
\langle \bbox{E}^2(\bbox{r})\rangle = \frac{C_2 \alpha_s a^2} {4 \pi r^6}
(1+3 \cos^2 \theta ),
\end{eqnarray}
where $C_2=(N^2-1)/2N$ is the Casimir eigenvalue for the fundamental
representation of the SU(N)$_{\rm color}$ group, $\alpha_s=g^2/4\pi$,
$g$ is the strong interaction coupling constant, and $\theta$ is the
angle between $\bbox{a}$ and $\bbox{r}$. On the other hand, the square
of the electric field strength in QED at a distance $r$, with $
a/\alpha > r > a$, in the vicinity of an electric charge dipole
represented by a Wilson loop, is \cite{Bha79b} (see also \cite{Luc91})
\begin{eqnarray}
\label{qed1}
\langle \bbox{E}_{\rm QED}^2(\bbox{r})\rangle = \frac{ \alpha a^2} 
{4\pi r^6} (1+3 \cos^2 \theta),
\end{eqnarray}
where $\alpha=e^2/4\pi$ is the fine-structure constant.  Therefore, if
the quark has an effective charge $c_q=\sqrt{C_2}$ (in unit of $g$)
and the antiquark an equal and opposite effective charge $c_{\bar
q}=-\sqrt{C_2}$ (in unit of $g$), and each effective charge generates
a color electric field strength as as in QED, then the square of the
total color electric field strength $\bbox{E}^2$ in the vicinity of a
meson will be the same as given in Eq.\ (\ref{ext}) obtained by
summing up two-gluon emission diagrams in QCD.  Because of the
similarities between Eqs.\ (\ref{ext}) and (\ref{qed1}), effective
charges in a QED-type interaction is a simple way to represent
multi-quark QCD interactions in the vicinity of a meson.  At short
distances, the effective charges in a QED-type interaction also leads
to the proper QCD color-Coulomb interaction between a quark and an
antiquark in a color-singlet state
\begin{eqnarray}
\label{int}
V_c(r)=- \frac {C_2 \alpha_s}{r}.
\end{eqnarray}

The interaction between a quark and an antiquark contains not only of
the color-Coulomb component but also the linear confining potential.
We need to modify the color-Coulomb interaction to include both the
color-Coulomb potential and the linear confining potential.  Recent
lattice gauge calculations show that the nonperturbative interquark
potential between a quark and an antiquark in different
representations is proportional to the eigenvalue of the quadratic
Casimir operator $C_2$ \cite{Bal00}.  Thus, there is the Casimir
scaling in the interaction of a quark and an antiquark and the concept
of the effective charge is also applicable when the potential is
generalized to include both the color-Coulomb and the confining
linear components.

Color interaction given by Eq.\ (\ref{ext}) is the well-known van der
Waals interactions arising from repeated applications of the one-gluon
exchange interaction for extended and separated mesons
\cite{App77,Pes79,Bha79a,Bha79b,Lip82,Gre81,others}.  There is however
no experimental evidence for the inverse-power color van der Waals
interaction between separated mesons.  This experimental
contradiction led Lipkin and Greenberg
\cite{Lip82} to note that the quark degree of freedom with only
one gluon-exchange interaction for a multi-quark system is incomplete.
The gluon in non-Abelian QCD with its color charges and nonlinear
self-interaction is very different from the neutral photon.  The
gluonic degree of freedom is needed in multi-quark interactions
\cite{Lip82}.  One can represent the gluonic degree of freedom in
terms of a string, and the important effect of the dynamical string is
the breaking of the string at large distances, resulting in the
screening of the long-range interaction.  Accordingly, we should
modify the phenomenological ``color-Coulomb plus linear'' interaction
further by introducing a screening mass $\mu$, because we intend to
use the effective interaction also for distances larger than 1 fm for
our molecular state.

When Eq.\ (\ref{int}) is written as $V_c(r)=-C_2 v_c(r)$ by separating
out the color factor $(-C_2)$, the basic color-Coulomb interaction is
$v_c(r)= \alpha_s/r$ and the corresponding basic linear confining
interaction is $v_{\rm lin}(r)= -3 br/4$.  They can be represented in
the momentum space by $\tilde v_c (\bbox{k})= 4 \pi
\alpha_s/\bbox{k}^2$ and $\tilde v_{\rm lin} (\bbox{k})= 6 \pi
b/\bbox{k}^4$.  The effect of screening can be introduced by replacing
$\bbox{k}^2$ with $\bbox{k}^2+\mu^2$, leading to an interaction in
momentum space given by \cite{Zha03}
\begin{eqnarray}
{\widetilde v}({\bf k})=\bigg[{{4\pi \alpha_s}\over{ {\bf k}^2 + \mu^2 }}
+ { {6\pi b} \over { ( {\bf k}^2 + \mu^2 )^2 } }
\bigg] \,.
\end{eqnarray}
From the interaction in momentum space, we can obtain the interaction
in the the configuration space. The screened color-Coulomb interaction
becomes the Yukawa interaction, and the screened linear potential
becomes the exponential interaction. One also needs to include the
hyperfine spin-spin interaction.  As the spin-spin interaction is
short-ranged, the effect of screening on the spin-spin interaction
can be neglected.

The effective interaction between quark and antiquark particles $j$
and $k$ in the four-body system is then given by
\begin{eqnarray}
\label{potena}
V_{jk}(\bbox{r}_{jk})= c_{j} c_{k} v(\bbox{r}_{jk})
\end{eqnarray}
where
\begin{eqnarray}
\label{poten0}
c_j=
\begin{cases}
{~~~\sqrt{C_2}}& {\rm if~}j{\rm~is~a~quark},\cr
{-\sqrt{C_2}}& {\rm if~}j{\rm~is~an~antiquark},\cr
\end{cases}
\end{eqnarray}
and
\begin{eqnarray}
\label{potenb}
v(\bbox{r}_{jk})=
 \frac{\alpha_s e^{-\mu {r}_{jk}}}{{r}_{jk}} 
+\frac {3b}{4\mu} {e^{-\mu {r}_{jk}}}
- \frac{8 \pi \alpha_s }{3 m_j m_k } \bbox{s}_j \cdot \bbox{s}_k \left
(\frac {\sigma^3}{ \pi^{3/2} }\right ) e^{-\sigma^2 r_{jk}^2}.
\end{eqnarray}
The central potential reduces to the usual Cornell-type potential in
the limit when $\mu \to 0$.  With this effective interaction, a quark
and an antiquark in the four-body molecular state $(Q \bar q)$-$(q\bar
Q)$ system are subject to the same phenomenological interaction and
can lead to the proper meson bound states.

The residue interaction between $A(12)$ and $B(34)$ is then
\begin{eqnarray}
\label{vi}
V_I=\sum_{j=1}^2 \sum_{k=3}^4 c_{j} c_{k} v(\bbox{r}_{jk}).
\end{eqnarray}

By making use of earlier studies of potential parameters and a running
coupling constant \cite{Won02}, we use an effective interaction with
the following set of parameters for our present work
\begin{eqnarray}
\label{par}
\alpha_s(Q^2)&=&{12 \pi \over (33-2n_f) \ln(A+Q^2/B^2)},
~~~A=10,~~~B=0.31 {\rm ~GeV},
 \nonumber\\
b&=&0.335 {\rm ~GeV}^2,~~~\mu=0.28 {\rm~GeV},~~~\sigma=0.897 {\rm~GeV},
\\
m_u&=&m_d=0.334 {\rm ~ GeV},~~m_c=1.87 {\rm
~GeV}, ~~m_b=5.18 {\rm~~GeV}.\nonumber
\end{eqnarray}
The string tension coefficient $b=0.335$ GeV$^2$ and the screening mass
$\mu=0.28$ GeV were previously found to give reasonable descriptions
of the charmonium masses \cite{Won99}.  The value of the screening
mass gives a screening length of $1/\mu\sim 0.7$ fm which is
consistent with the minimum string-breaking distance $L_{\rm min}$,
(Eq.\ (5.11) of \cite{Won94}), based on the Schwinger mechanism
\cite{Sch51} of pair production,
\begin{eqnarray}
L_{min}=2m_{u,d}/\kappa=0.67 {\rm fm},
\end{eqnarray}
where we have used a string tension coefficient $\kappa$ of 1 GeV/fm.
The large value of $b$ in Eq.\ (\ref{par}) arises because the
effective string tension coefficient $be^{- \mu r_{jk}}$ is not
constant and it varies with $r_{jk}$. In Eq.\ (\ref{par}), we identify
$Q$ as the sum of the rest masses of the interacting quarks.  The
parameterization of the phenomenological strong coupling constant
$\alpha_s$ and the spin-spin interaction parameter $\sigma$ come from
earlier calculations of meson masses in
\cite{Won02}.  
The set of parameters in Eq.\ (\ref{par}) leads to a good
description of the masses of $D$, $D^*$, $J/\psi$, $B$, $B^*$, and
$\Upsilon$ mesons.

\section{Evaluation of the meson-meson potential $V(\bbox{r})$}

We represent the wave function for the coordinate of the quark $Q$
relative to the antiquark $\bar q$ in the meson $A(12)$ by a Gaussian
wave function of the form
\begin{eqnarray}
\label{phiwf}
\phi_{lm}^{A}(\bbox{r}_{12})=\sqrt{ \frac{4\pi} 
{(2l+1)!!}}\left ( \frac{\beta_{A}^2}{\pi}\right )^{3/4}
(2\beta_{A}^2)^{l/2} i^l r_{12}^l e^{-\beta_{A}^2 r_{12} ^2/2}
Y_{lm}(\theta_{12},\phi_{12}).
\end{eqnarray}
The wave function is normalized according to
\begin{eqnarray}
\int 
d\bbox{r} |\phi_{lm}^{A}(\bbox{r}_{12})|^2=1.
\end{eqnarray}
A similar wave function $\phi_{l'm'}^{B}(\bbox{r}_{34})$ can be
written for the relative motion between the quark $q$ and the
antiquark $\bar Q$ in the meson $B(34)$.  

With the knowledge of the Gaussian wave functions and the effective
interquark interaction, the meson-meson potential $V(r)$ can be
evaluated analytically. The meson-meson potential depends on $\langle
A_\lambda B_{\lambda'}| V_I | A_0 B_{0}\rangle$ which can be further
decomposed as a sum of $\langle A_\lambda B_{\lambda'}|
v(\bbox{r}_{jk}) | A_0 B_{0}\rangle$,
\begin{eqnarray}
\langle A_\lambda B_{\lambda'}| V_I | A_0 B_{0}\rangle
=\sum_{j=1}^2 \sum_{k=3}^4
c_j c_k \langle A_\lambda B_{\lambda'}| v(\bbox{r}_{jk}) 
| A_0 B_{0}\rangle.
\end{eqnarray}
The matrix elements $\langle A_\lambda B_{\lambda'}| v(\bbox{r}_{jk})
| A_0 B_{0}\rangle$ can be evaluated in the configuration space where
there are three independent coordinates because the coordinates of the
four particles are connected by the center-of-mass coordinate
$\bbox{R}_{1234}$.  We can fix the center-of-mass coordinate
$\bbox{R}_{1234}$ to be at the origin,
\begin{eqnarray}
\label{e5}
\bbox{R}_{1234}=
 (m_{12}\bbox{R}_{12}
+m_{34}\bbox{R}_{34})/(m_{12}+m_{34})=\bbox{0},
\end{eqnarray}
and choose the independent coordinates to be $\bbox{r},\bbox{r}_{12}$,
and $\bbox{r}_{34}$.  Using Eqs.\ (\ref{rjk}), (\ref{e2}), (\ref{e3}),
(\ref{e4}), and (\ref{e5}), we can represent all other coordinates in
terms of these three coordinates:
\begin{eqnarray}
\label{rrr}
\bbox{r}_{jk}=\bbox{r}+f_A(jk)\bbox{r}_{12}+f_B(jk)\bbox{r}_{34},
\end{eqnarray}
where
\begin{eqnarray}
\label{rrr1}
f_A(14)=~~f_2, && f_B(14)=~~f_3,\nonumber\\ f_A(13)=~~f_2, &&
f_B(13)=-f_4, \\ f_A(23)=-f_1, && f_B(23)=-f_4,\nonumber\\
f_A(24)=-f_1, && f_B(24)=~~ f_3\nonumber.
\end{eqnarray}
Using the above relations between $\bbox{r}_{jk}$ and $\{\bbox{r},
\bbox{r}_{12},\bbox{r}_{34}\}$, the matrix element $\langle A_\lambda
B_{\lambda'}| v(\bbox{r}_{jk}) | A_0 B_{0}\rangle$ can be written as
\begin{eqnarray}
\label{mat}
\langle A_\lambda B_{\lambda'}| v(\bbox{r}_{jk}) |
A_0 B_{0}\rangle
=\int d\bbox{r}_{12} \int d\bbox{r}_{34}
\rho_{\lambda 0}^A (\bbox{r}_{12}) \rho_{\lambda' 0}^B (\bbox{r}_{34})
v(\bbox{r}+f_A(jk)\bbox{r}_{12}+f_B(jk)\bbox{r}_{34}),
\end{eqnarray}
where $\rho_{\lambda 0}^A$ and $\rho_{\lambda' 0}^B$ are
\begin{eqnarray}
\rho_{\lambda 0}^A(\bbox{r}_{12})
= \phi_{\lambda}^*(\bbox{r}_{12}) \phi_0(\bbox{r}_{12}),
\end{eqnarray} 
\begin{eqnarray}
\rho_{\lambda' 0}^A(\bbox{r}_{34})
= \phi_{\lambda'}^*(\bbox{r}_{34}) \phi_0(\bbox{r}_{34}).
\end{eqnarray} 
We note that $\rho_{0 0}^{A}(\bbox{r}_{12})$ is the density
distribution of meson $A_0 (12)$ and $\rho_{\lambda 0}^A
(\bbox{r}_{12})$ is the transition density for making the transition
from the ground state $A_0(12)$ to the excited states $A_\lambda(12)$.
Similarly, $\rho_{0 0}^{B}(\bbox{r}_{34})$ and $\rho_{\lambda' 0}^B
(\bbox{r}_{12})$ are respectively the density distribution and
transition density of meson $B(12)$.  By the method of Fourier
transform \cite{Sat84,Pet75}, the evaluation of the matrix element
(\ref{mat}) can be greatly simplified.  The result is
\begin{eqnarray}
\langle A_\lambda B_{\lambda'}| v(\bbox{r}_{jk}) |
A_0 B_{0}\rangle
=\int \frac{d\bbox{p}}{(2\pi)^3}
e^{i \bbox{p}\cdot \bbox{r}}
~{\tilde \rho}_{\lambda 0}^A[f_A(jk)\bbox{p}]
~~{\tilde \rho}_{\lambda' 0}^B[f_B(jk)\bbox{p}]
~~{\tilde v}(\bbox{p})
\end{eqnarray}
where
\begin{eqnarray}
{\tilde \rho}_{\lambda 0}^{A,B}(\bbox{p})
=\int {d\bbox{y}} e^{i \bbox{p}\cdot \bbox{y}}
\rho_{\lambda 0}^{A,B}(\bbox{y}),
\end{eqnarray}
\begin{eqnarray}
{\tilde v}(\bbox{p})
=\int {d\bbox{r}_{jk}} e^{-i \bbox{p}\cdot \bbox{r}_{jk}}
 v (\bbox{r}_{jk}).
\end{eqnarray}
We first evaluate the matrix element $\langle A_0 B_0|
v(\bbox{r}_{jk}) | A_0 B_0\rangle$ which is also needed in subsequent
calculations. From the wave functions we can determine
\begin{eqnarray}
{\tilde \rho}_{00}^{A,B}(\bbox{p})
=e^{-\bbox{p}^2/4 \beta_{A,B}^2}.
\end{eqnarray}
Therefore, the matrix element $\langle A_0 B_0| v(\bbox{r}_{jk}) | A_0
B_0\rangle$ is related to ${\tilde v}(\bbox{p})$ by
\begin{eqnarray}
\label{00}
\langle A_0 B_0| v(\bbox{r}_{jk}) | A_0 B_0\rangle
=\int \frac{p^2 dp\, d\mu\, d\phi}{(2\pi)^3} e^{i \bbox{p} \cdot \bbox{r}}
e^{-a_{jk}^2 p^2} {\tilde v} (p) \equiv v_{0 0}(jk; a_{jk}^2,{r}),
\end{eqnarray}
where $a_{jk}^2$ is
\begin{eqnarray}
a_{jk}^2= \frac {[f_A(jk)]^2}{4 \beta_A^2} 
+ \frac {[f_B(jk)]^2}{4 \beta_B^2},
\end{eqnarray}
and we have written $\langle A_0 B_0| v(\bbox{r}_{jk}) | A_0
B_0\rangle$ as $v_{0 0}(jk; a_{jk}^2,{r})$ to exhibit fully its
dependence on $a_{jk}^2$ and $r$.

The matrix element $\langle A_0 B_0| v(\bbox{r}_{jk}) | A_0
B_0\rangle$ can be expressed as the sum of contributions from the
Yukawa interaction and the exponential interaction in Eq.\
(\ref{potenb}),
\begin{eqnarray}
\label{v01}
 \langle A_0 B_0| v(\bbox{r}_{jk}) | A_0 B_0\rangle
=\langle A_0 B_0| v^{\rm Yuk}(\bbox{r}_{jk}) | A_0 B_0\rangle
+\langle A_0 B_0| v^{\rm exp}(\bbox{r}_{jk}) | A_0 B_0\rangle.
\end{eqnarray}
The matrix element for the Yukawa interaction is found to be
\begin{eqnarray}
\label{v02}
 \langle A_0 B_0| v^{\rm Yuk}(\bbox{r}_{jk}) | A_0 B_0\rangle
= \alpha_s e^{a_{jk}^2\mu^2}(u_1 v_1-u_2 v_2),
\end{eqnarray}
\begin{eqnarray}
u_1= \left ( 1+{\rm erf} [(r-2 a_{jk}^2 \mu)/(2 a_{jk})] \right )/2,
\end{eqnarray}
\begin{eqnarray}
u_2= \left ( 1-{\rm erf} [(r+2 a_{jk}^2 \mu)/(2 a_{jk})] \right )/2,
\end{eqnarray}
\begin{eqnarray}
v_1= \frac {e^{ -\mu r}}{r},
\end{eqnarray}
\begin{eqnarray}
v_2= \frac {e^{ {}\mu r}}{r}.
\end{eqnarray}
The matrix element for the exponential interaction is
\begin{eqnarray}
\label{v03}
 \langle A_0 B_0| v^{\rm exp}(\bbox{r}_{jk}) | A_0 B_0\rangle
= \frac {3b}{4\mu} e^{a_{jk}^2\mu^2}(u_1 w_1+u_2 w_2),
\end{eqnarray}
where
\begin{eqnarray}
w_1= (r-2a_{jk}^2 \mu) \frac {e^{ -\mu r}}{r},
\end{eqnarray}
and
\begin{eqnarray}
w_2= (r+2 a_{jk}^2 \mu) \frac {e^{ {}\mu r}}{r}.
\end{eqnarray}
The direct potential vanishes when all the quark masses are equal,
and it becomes more attractive as the mass of the heavy quark
increases.

We fix the vector $\bbox{r}$ to lie along the $z$-axis, and quantize
the azimuthal component of the angular momentum $m$ to be projections
along the $z$-axis.  For the polarization potential $V_{\rm pol}(r)$,
we consider intermediate excitations to the next excited states of
$A(12)$ and $B(12)$ characterized by $\{\lambda
{\rm~or~}\lambda'=lm=1m\}$.  It can be shown that the matrix element
$\langle A_\lambda B_{\lambda'}| v(\bbox{r}_{jk}) | A_0 B_{0}\rangle$
is zero for $\{\lambda=0; \lambda'=1m\}$ and
$\{\lambda=1m;\lambda'=0\}$.  It gives a non-vanishing matrix element
for $\{\lambda=1m; \lambda'=1-m\}$ with $\{m=-1,0,1\}$ when both
mesons are excited.

To evaluate the matrix element $\langle A_{1m}
B_{1-m}|v(\bbox{r}_{jk}) | A_0 B_0\rangle$, we write out the wave
functions explicitly and obtain
\begin{eqnarray}
\langle A_{10} B_{10}| v(\bbox{r}_{jk}) | A_0 B_0\rangle
=\frac{f_A(jk) f_B(jk)}{2 \beta_A \beta_B}
\int \frac{p^2 dp\, d\mu\, d\phi}{(2\pi)^2} e^{i \bbox{p} \cdot \bbox{r}}
e^{-a_{jk}^2 p^2} p^2 \mu^2 {\tilde v} (p).
\end{eqnarray}
A comparison of the above equation with Eq.\ (\ref{00}) yields
\begin{eqnarray}
\langle A_{10} B_{1 0}| v(\bbox{r}_{jk}) | A_0 B_0\rangle
=-\frac{f_A(jk) f_B(jk)}{2 \beta_A \beta_B}
  \frac{\partial^2}{\partial r^2} 
 v_{0 0}(jk; a_{jk}^2,{r}).
\end{eqnarray}
To evaluate the matrix element $\langle A_{11} B_{1 -1}|
v(\bbox{r}_{jk}) | A_0 B_0\rangle$, we write out the wave functions
explicitly and obtain
\begin{eqnarray}
\langle A_{11} B_{1-1}| v(\bbox{r}_{jk}) | A_0 B_0\rangle
=\frac{f_A(jk) f_B(jk)}{2 \beta_A \beta_B}
\int \frac{p^2 dp\, d\mu\, d\phi}{(2\pi)^3} e^{i \bbox{p} \cdot \bbox{r}}
e^{-a_{jk}^2 p^2}  \frac{p^2(1-\mu^2)}{2}  {\tilde v} (p).
\end{eqnarray}
A comparison of the above equation with Eq.\ (\ref{00}) gives
\begin{eqnarray}
\langle A_{11} B_{1 -1}| v(\bbox{r}_{jk}) | A_0 B_0\rangle
=\frac{f_A(jk) f_B(jk)}{2 \beta_A \beta_B}
\frac {1} {2} \left ( - \frac{\partial}{\partial a_{jk}^2}
                      + \frac{\partial^2}{\partial r^2} \right )
 v_{0 0}(jk; a_{jk}^2,{r}).
\end{eqnarray}
With the analytical results of Eqs.\ (\ref{v01}), (\ref{v02}), and
(\ref{v03}) for $\langle A_0 B_0| v(\bbox{r}_{jk}) | A_0 B_0\rangle$
(or $v_{0 0}(jk; a_{jk}^2,{r})$), the differentiation of $v_{00}(jk;
a_{jk}^2,{r})$ with respect to $a_{jk}^2$ and $r$ can be readily
carried out analytically to give the matrix elements $\langle A_{1m}
B_{1 -m}| v(\bbox{r}_{jk}) | A_0 B_0\rangle$ for $\{m=-1,0,1\}$.  The
energy denominator in Eq.\ (\ref{polar}),
$\{\epsilon(A_{1m})+\epsilon(B_{1-m}) -\epsilon(A_0)-\epsilon(B_0)\}$,
can be obtained from experimental masses of $D(1P)$, $\bar D^*(1P)$,
$D(1S)$, and $\bar D^* (1S)$.  The masses of the 1P states of the open
bottom mesons are not known experimentally.  We infer the mass
differences from theoretical masses of open bottom mesons calculated
in \cite{Won02} where we get $M(B(1P))-M(B_0)=0.5348$ GeV, and
$M(B^*(1P))-M(B_0^*)=0.5168$ GeV.  These quantities allows us to
calculate the polarization potential using Eq.\ (\ref{polar}).

\section{Heavy Quark Molecular States}

\begin{figure}[h]
\includegraphics[scale=0.6]{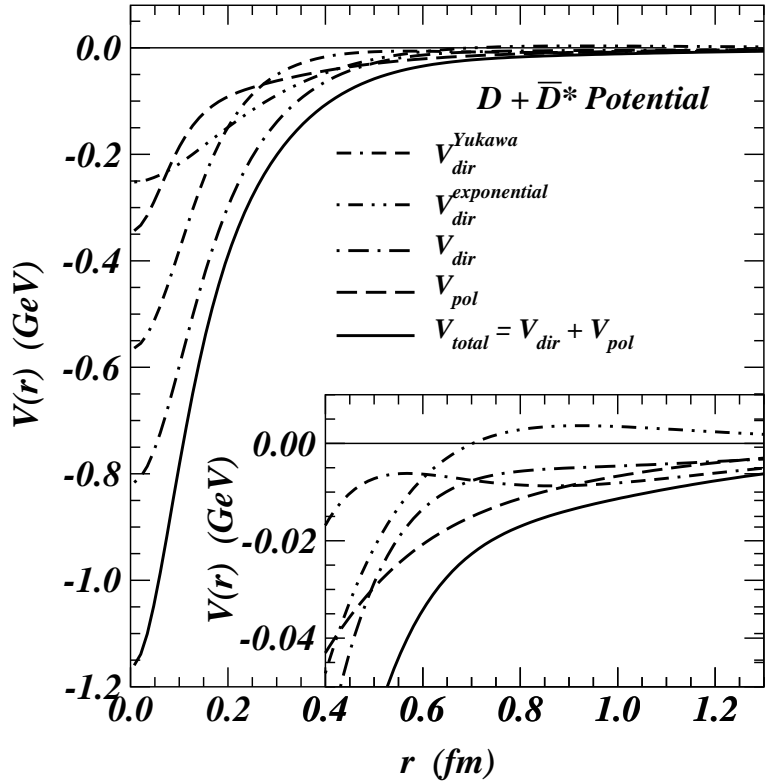}
\caption{ The total potential $V(r)$ and its various components for
the $D$$\bar D^*$ system.  The direct potential $V_{\rm dir}$ is
the sum of $V_{\rm dir}^{\rm Yukawa}$ and $V_{\rm dir}^{\rm
exponential}$, and $V(r)$ is the sum of the direct potential and the
polarization potential. The insert with an expanded ordinate scale gives a
clearer picture of the total potential and its various components at
large distances.  }
\end{figure}

The meson wave function in Eq.\ (\ref{phiwf}) is characterized by a
momentum width parameter $\beta$ that is related to the
root-mean-square quark-antiquark separation $r_{\rm rms}$ of the meson
by
\begin{eqnarray}
\beta=\sqrt{\frac{3}{2}}~ \frac{1} {r_{\rm rms}}.
\end{eqnarray}
Previously, root-mean-square quark-antiquark separation of mesons have
been obtained from calculations of meson wave functions and meson
masses (Table IV of \cite{Won02}). From this table, we use the
root-mean-square quark-antiquark separations of $r_{\rm rms}(D)=0.585$
fm, $r_{\rm rms}(D^*)=0.626$ fm, $r_{\rm rms}(B)=0.574$ fm, and
$r_{\rm rms}(D^*)=0.583$ fm to calculate the $\beta$ values for
various meson wave functions. In the present numerical work, we shall
include only the screened color-Coulomb interaction (Yukawa
interaction) and the screened linear interaction (exponential
interaction) so as to study the gross features of the meson-meson
potential and molecular states.

\begin{figure}[h]
\includegraphics[scale=0.6]{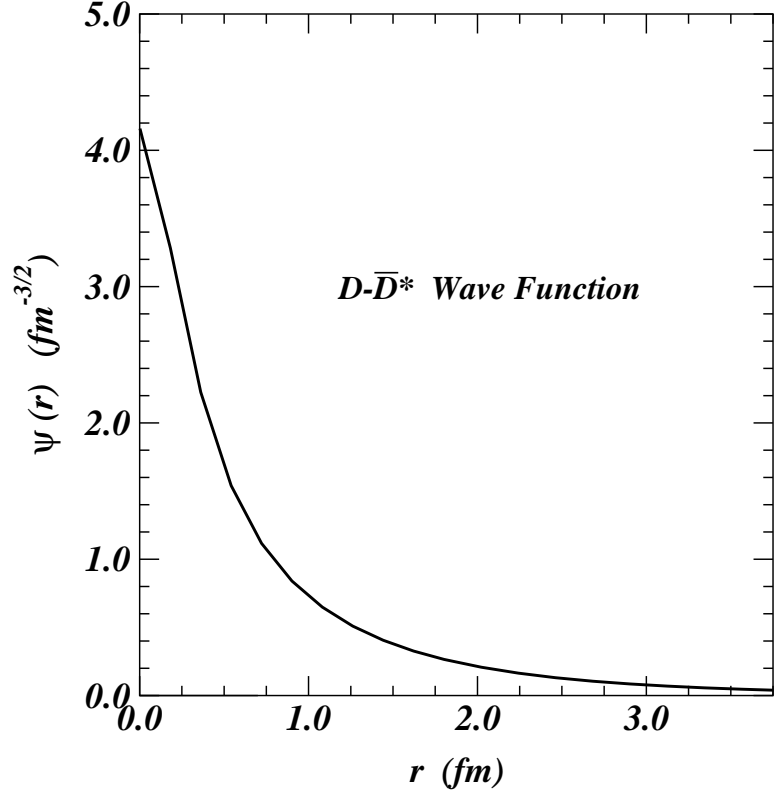}
\caption{ The wave function $\psi(r)$ of the $D$$\bar D^*$ system
as a function of the separation between $D$ and $\bar D^*$.  }
\end{figure}

Using the effective interaction outlined above, we obtained
meson-meson potentials for various combinations of $D, D^*, B, B^*$
with $\bar D, \bar D^*, \bar B, \bar B^*$.  The total potential $V(r)$
and its various components for $D$$\bar D^*$ are shown in Fig.\ 3.
The insert in Fig. 3 gives a more expanded description of various
components at large distances.  The total direct potential is the sum
of the contributions from the Yukawa and the exponential interaction,
and is attractive. The magnitude of the polarization potential is
smaller than the magnitude of the direct potential at short distances
but the role is reversed at intermediate separations.  The two
different contributions are about equal at large separations beyond
1.2 fm.  As a result, the total potential, which is the sum of the
direct potential and the polarization potential, has an extended
attractive region extending to 1 fm and beyond.

With such a potential and the experimental masses of $D$ and $\bar
D^*$, we solve for the lowest eigenstate of the meson-meson system. We
find a bound state at a binding energy of 7.51 MeV.  We plot the wave
function for this molecular state in Fig. 4.  The wave function
$\psi(r)$ is quite extended and has a significant amplitude beyond 1
fm.  This state is weakly bound, and has a large root-mean-square
$D$-$\bar D^*$ separation of 1.37 fm.  Its weak binding and its
extended spatial separation between $D$ and $\bar D^*$ provide a good
characterization of its molecular structure.

For a flavor-independent effective interaction, the theoretical
meson-meson potentials and the binding energies are expected to change
very insignificantly for different flavors of the light quarks.  On
the other hand, the mass of the system is related to the binding
energy $B$ by
\begin{eqnarray}
M=M(A_\nu)+M(B_{\nu'})+\epsilon=
M(A_\nu)+M(B_{\nu'})-B,
\end{eqnarray}
and the masses of $A_\nu$ and $B_{\nu'}$ depend on light quark
flavors.  Therefore, the calculated mass of the $(Q\bar q)$-$(q\bar
Q)$ molecular system depends on light quark flavors.  We list the
molecular states for different combinations of the $(c\bar q)$ and
$(q\bar c)$ mesons in Table I.  The properties of the states in the
$D$$\bar D$, $D$$\bar D^*$, and $D^*$$\bar D^*$ systems are quite
similar.

\vskip 0.5cm {Table I.  The calculated masses, binding energies, and
the root-mean-square $(c\bar q)$-$(q\bar c)$ separation $r_{\rm rms}$
of molecular states formed by $(c\bar q)$ and $(q\bar c)$ mesons.}

\vspace*{0.3cm}\hspace*{1.8cm}
\begin{tabular}{|c|c|c|c|c|} 
\hline
\vspace*{-0.3cm}
     & & & & \\
~~$ (c\bar q)$-$ (q\bar c)$~~~  & ~~State~~  &
 Binding Energy (MeV)  & ~~~Mass~(MeV) ~~~& $r_{\rm rms}$ (fm)    \\
\hline
\vspace*{-0.3cm}
     & & & & \\
  $D^0$$\bar D^0$                    & 1S  &  3.10   &  3725.90 &  1.73  \\  
  $D^0$$D^-$,  $D^+$$\bar D^0$       & 1S  &  3.10   &  3730.70 &  1.73  \\  
  $D^+$$D^-$                         & 1S  &  3.10   &  3735.50 &  1.73  \\  
\hline
\vspace*{-0.3cm}
     & & & & \\
  $D^0$$\bar D^{*0}$                 & 1S   &  7.53  &  3863.67 &  1.37  \\
  $D^0$$D^{*-}$                      & 1S   &  7.53  &  3866.97 &  1.37  \\
  $D^+$$\bar D^{*0}$                 & 1S   &  7.53  &  3868.47 &  1.37  \\
  $D^+$$D^{*-}$                      & 1S   &  7.53  &  3871.77 &  1.37  \\
\hline
\vspace*{-0.3cm}
     & & & & \\
  $D^{*0}$$\bar D^{*0}$              & 1S   & 15.31  &  3998.09 &  1.06   \\
  $D^{*0}$$D^{*-}$,$D^{*+}$$\bar D^{*0}$ & 1S & 15.31 & 4001.39 & 1.06   \\
  $D^{*+}$$D^{*-}$                   & 1S   & 15.31  &  4004.69 &  1.06   \\
\hline
\end{tabular}

\vskip 0.5cm To produce the unknown state $X(3872)$ in the process
$B^{\pm} \to K^{\pm} X(3872)$ observed by the Belle Collaboration, the
state $X(3872)$ most likely originates from the production of a $c\bar
c$ pair with $I=0$ during the weak decay of the bottom quark.  On the
other hand, the calculated masses of $D^0$$\bar D^{*0}$ and
$D^+$$D^{*-}$ molecular systems are respectively 3863.67 MeV and
3871.77 MeV (see Table I).  By the symmetry of charge conjugation, the
masses of $D^{*0}$$\bar D^{0}$ and $D^{-}$$D^{*+}$ systems are also
3863.67 MeV and 3871.77 MeV respectively.  The most suitable candidate
for the 3872 MeV state appears to be the molecular state $D^+ D^{*-}$
and its charge conjugate $D^{-}$$D^{*+}$. One expects that the
molecular states $D^0$$\bar D^{*0}$, $D^{+}$$\bar D^{*-}$, and
$D^{-}$$D^{*+}$ will be mixed to form components of $I=0$ and $I=1$
states.  As they lie in the vicinity of the 3872 state, it is tempting
to identify the $I=0$ of this set of multiplets as the 3872 state
observed by the Belle Collaboration. It has been conjectured however
that the 3872 state may break isospin symmetry maximally
\cite{Clo03,Vol03} and more experimental information are needed to
understand better the isospin properties of the 3872 state.

\begin{figure}[h]
\includegraphics[scale=0.6]{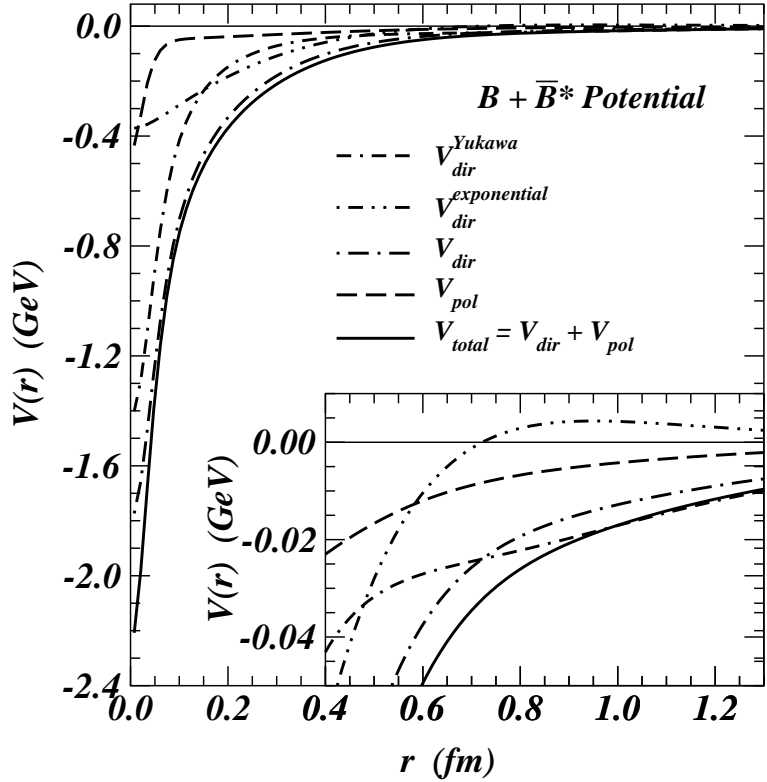}
\caption{ The total potential $V(r)$ between $B$ and $\bar B^*$ and
its various components. The insert gives the total potential and its
various components at large separations in greater detail.}
\end{figure}

\vspace*{0.3cm} We show in Fig. 5 the potential for the $B$$\bar B^*$
system. The potential has features similar to those of the $D$$\bar
D^*$ system except that the $B$$\bar B^*$ potential is deeper than the
$D$$\bar D^*$ potential.  This arises because the quark mass ratio
$m_b/m_{u,d}$ is greater than $m_c/m_{u,d}$.  For the $B\bar B^*$
system, the magnitude of the direct potential is consistently much
larger than the magnitude of the polarization potential.  The total
potential is attractive at large separations.  We search for
eigenstates for the $B$$\bar B^*$ system.  The large mass of the
bottom quark and the large depth of the $B$$\bar B^*$ potential leads
to two eigenstates of the $B\bar B^*$ system.  There is a 1S state
with a binding energy of 151.75 MeV and a root-mean-square $B$-$\bar
B^*$ separation of 0.314 fm.  In addition, there is a $2S$ state with
a binding energy of 0.88 MeV and a root-mean-square $B$-$\bar B^*$
separation of 2.23 fm.

The wave functions for these two states are shown in Fig. 6.  The
$\psi_{1S}(r)$ wave function is confined at short distances, but the
$\psi_{2S}(r)$ wave function is spatially quite extended.  We list in
Table II the calculated binding energies, masses, and the
root-mean-square radii of states formed by $ (b\bar q)$ and $ (q\bar
b)$.  As the masses of the $B^0$ and $B^{\pm}$ are nearly the same, we
do not need to distinguish $B$ (and similarly $B^*$) mesons of
different light quark flavors.

\vskip 0.5cm {Table II.  The calculated masses, binding energies, and
the root-mean-square $(b\bar q)$-$(q\bar b)$ separation $r_{\rm rms}$
of molecular states formed by $(b\bar q)$ and $(q\bar b)$ mesons. }

\vspace*{0.3cm}\hspace*{1.8cm}
\begin{tabular}{|c|c|c|c|c|} 
\hline
\vspace*{-0.3cm}
     & & & & \\
~~$ (b\bar q)$-$ (q\bar b)$~~~  & ~~State~~  &
 Binding Energy (MeV)  & ~~~Mass~(MeV) ~~~& $r_{\rm rms}$ (fm)    \\
\hline
\vspace*{-0.3cm}
     & & & & \\
   $B$$\bar B$      & 1S   &  148.24  & 10410.56    &  0.317  \\  
  $B$$\bar B^*$    & 1S   &  151.75  & 10452.64    &  0.314  \\
  $B^*$$\bar B^*$  & 1S   &  155.40  & 10494.60    &  0.311  \\
  $B$$\bar B$      & 2S   &    0.80  & 10558.00    &  2.25   \\
  $B$$\bar B^*$    & 2S   &    0.88  & 10603.52    &  2.23   \\
  $B^*$$\bar B^*$  & 2S   &    0.95  & 10649.05    &  2.21   \\
\hline
\end{tabular}

\begin{figure}[h]
\includegraphics[scale=0.6]{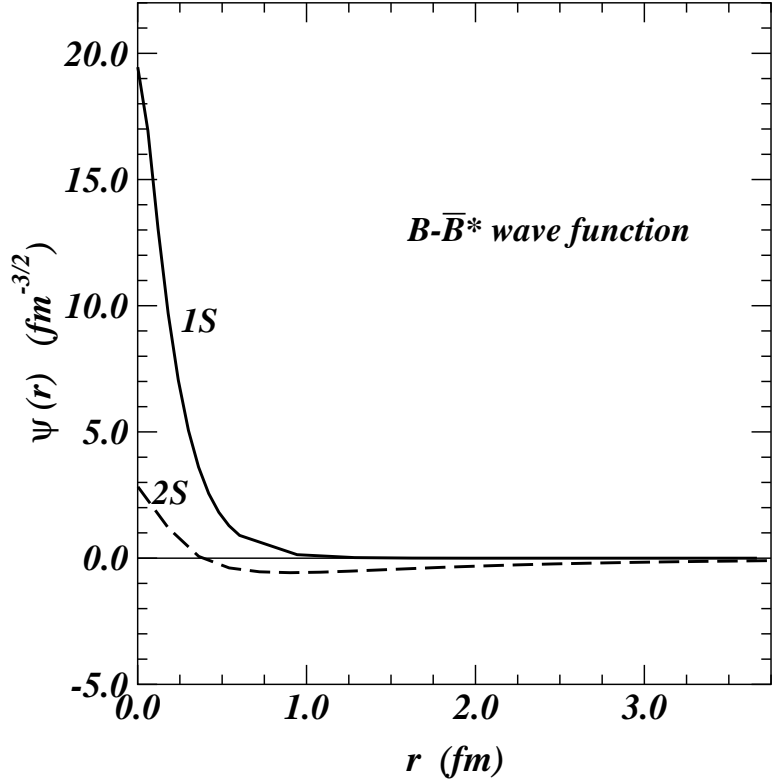}
\caption{ The wave function $\psi(r)$ of the $B$$\bar B^*$ system
as a function of the separation between $B$ and $\bar B^*$. 
}
\end{figure}

\vspace*{0.4cm}
The potentials and wave functions for $(c\bar q)$-$(q \bar b)$ systems
involving both charm quark and bottom quarks are similar to those of
$B$$\bar B^*$ and $D$$\bar D^*$ and have similar $1S$ eigenstates with
a binding energy about 29-43 MeV and a root-mean-square $(c\bar
q)$-$(q\bar b)$ separation about 0.6-0.8 fm.  We list in Table III the
eigenstates for various $(c\bar q)$-$(q\bar b)$ systems.

\vskip 0.5cm {Table III.  The calculated masses, binding energies, and
the root-mean-square $(c\bar q)$-$(q\bar b)$ separation $r_{\rm rms}$
of molecular states formed by $(c\bar q)$ and $(q\bar b)$ mesons. }

\vspace*{0.3cm}\hspace*{1.8cm}
\begin{tabular}{|c|c|c|c|c|} 
\hline
\vspace*{-0.3cm}
     & & & & \\
~~$ (c\bar q)$-$ (q\bar b)$~~~  & ~~State~~  &
 Binding Energy (MeV)  & ~~~Mass~(MeV) ~~~& $r_{\rm rms}$ (fm)    \\
\hline
\vspace*{-0.3cm}
     & & & & \\
  $D^0$$\bar B$    & 1S   &   27.83  &  7115.01    &  0.756  \\  
  $D^+$$\bar B$    & 1S   &   27.83  &  7119.81    &  0.756  \\  
\hline
\vspace*{-0.3cm}
     & & & & \\
  $D^0$$\bar B^*$  & 1S   &   29.36  &  7158.96    &  0.739  \\  
  $D^+$$\bar B^*$  & 1S   &   29.36  &  7163.76    &  0.739  \\  
\hline
\vspace*{-0.3cm}
     & & & & \\
  $D^{*0}$$\bar B$ & 1S   &   40.48  &  7244.07    &  0.638  \\  
  $D^{*+}$$\bar B$ & 1S   &   40.48  &  7247.38    &  0.638  \\  
\hline
\vspace*{-0.3cm}
     & & & & \\
  $D^{*0}$$\bar B^*$  & 1S   &   42.63  &  7287.34    &  0.624  \\  
  $D^{*+}$$\bar B^*$  & 1S   &   42.63  &  7290.64    &  0.624  \\  
\hline
\end{tabular}

\vskip 0.6cm

\vskip 0.6cm

\section{Potential for the $(Q\bar q)$-$(Q\bar q)$ system}

From our present model of effective charges, it is easy to see how the
direct potential for the $(Q\bar q)$-$(q\bar Q)$ system depends on
the ratio of the heavy quark mass $m_Q$ to the light quark mass $m_q$.
If the mass of $Q$ is the same as the mass of $q$, then the attraction
between effective charges of opposite signs will be compensated by the
repulsive interaction between effective charges of the same sign, when
we bring the two mesons together.  The net direct potential is zero
when $m_Q=m_q$.  The polarization potential is always attractive, but
the magnitude of the polarization potential in most cases is not
attractive enough to lead to a bound meson-meson system.

\begin{figure}[h]
\includegraphics[scale=0.6]{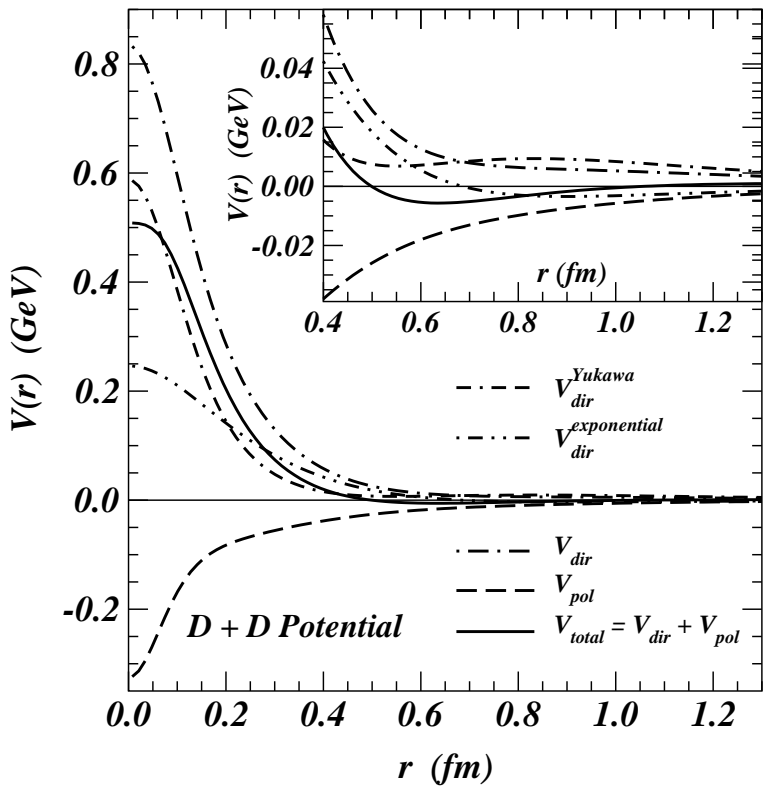}
\caption{ The total potential $V(r)$ between $D$ and $D$
is shown as the solid curve.  Shown also are various components of the
potential.  The insert gives the potentials at large separations in
greater detail.}
\end{figure}

If $m_Q \gg m_q$, the light quark and antiquark in $(Q\bar q)$-$(q\bar
Q)$ will move around a pair of nearly static heavy quark and heavy
antiquark.  As one brings a $(Q\bar q)$ in the vicinity of a $(q\bar
Q)$ meson, the light quark and the light antiquark will attract each
other because of their opposite effective charges and the meson-meson
interaction will be attractive.  The greater the mass ratio $m_Q/m_q$,
the greater their attraction will be.  Furthermore, the attractive
polarization potential extends to large separations and facilitates
the formation of molecular states.

From the above argument, one notes that the direct potential will be
repulsive for the $(Q\bar q)$-$(Q \bar q)$ system which has two heavy
quarks and two light antiquarks.  In this case, as the mesons are
brought close together, the two antiquarks moving around the nearly
static heavy quarks will interact. As the effective charges of the
antiquarks have the same sign, they tend to repel each other and the
resultant direct interaction is repulsive.  We show in Fig. 7 the
total potential and its various components for the $D$$D$ system which
has a repulsive direct potential.  Although the polarization
potential is attractive, the total potential is only very weakly
attractive between 0.5 fm and 1 fm with a maximum depth of about 6
MeV.  Such a shallow potential is not attractive enough to hold a
molecular state.  It should be pointed out, however, that
symmetrization or anti-symmetrization of the spatial wave functions of
the light antiquarks orbiting around the heavy quarks will give rise
to additional effects \cite{Mic99,Bar99}.  These effects must be
studied in more detail for the $QQ\bar q \bar q$ system, as in the
case of electrons in a diatomic molecule.  In these future studies,
the direct potential and the polarization potential will provide some
of the information needed for such an investigation.

\section{Conclusion and Discussions}

We have studied four-quark $(Q\bar q)$-$(q\bar Q)$ systems involving
charm and bottom quarks in a quark-based model.  The system is
described by a non-relativistic four-body Hamiltonian with a pairwise
interaction.  

We note that the molecular state of our interest is weakly-bound and
the average separation $R$ between the heavy quark mesons is
considerably greater than the average radius $a$ of the mesons.  As $R
>> a$, the dominant process for the interaction of the mesons is the
exchange of two gluons, which leads to the color van der Waals
interaction \cite{Pes79,Bha79a,Bha79b}.  We can equivalently represent
the color van der Waals interaction in terms of effective charges for
quarks and antiquarks in a QED-type interaction. These effective
charges give the proper description of bound states for a quark
interacting with an antiquark in an isolated mesons.  They also gives
rise to a color-electric dipole-dipole interaction between mesons when
a meson is brought in close proximity to another meson
\cite{Pes79,Bha79a,Bha79b}.  The color van der Waals interaction
should be screened by introducing a screening mass $\mu$ due to the
breaking of the gluonic string at large distance. Incorporating these
concepts in a phenomenological interquark potential, we find a
weakly-bound $D^+ D^{*-}$ state near the threshold that may be
qualitatively identified as the 3872 state observed recently by the
Belle Collaboration.

In the present model, the meson-meson potential for the $(Q\bar
q)$-$(q\bar Q)$ system can be decomposed into a direct potential and
a polarization potential.  The direct potential depends on the ratio
of the quark masses $m_Q/m_q$.  If the mass ratio is 1, then the
direct potential is zero.  The magnitude of the attractive direct
potential increases as the mass ratio increases.  The polarization
potential is always attractive.  However, only when the quark mass
ratio of both mesons are sufficiently high can the total potential be
attractive enough to lead to bound states.  There are no bound
molecular states for $(s\bar q)$-$(q\bar s)$ and $(c\bar q)$-$(q\bar
s)$ systems if we use meson-meson potentials obtained in the present
model.  We find bound meson-meson states in open heavy-quark systems
involving combinations of $\{ D, D^*, B, B^*\}$ with $\{\bar D, \bar
D^*, \bar B, \bar B^*\}$.  For open bottom meson pairs, we found
weakly-bound 2S states near the threshold, in addition to deeper 1S
states with binding energies of about 150 MeV.

We have included only screened color-Coulomb and screened linear
confining interactions in the present work. It will be of interest to
study in future work the influence of the spin-spin and other
components of the interaction on the molecular structure. In meson
systems having spin zero and one, the spin-spin interaction does not
contribute to the direct potential when the spin of one of the mesons
is zero.  The effect of spin-spin interaction is expected to be small
when one of the mesons is $D$, $B$, $\bar D$, or $\bar B$.

The present model differs from those of the pion-exchange models and
other previous multi-quark models
\cite{Tor03,Clo03,Pak03,Vol03,Bib03,Eic03,Cha03,Ruj77,Jaf76,Dol74,Liu80,Won80,Wei82,Tor92,Man92,Eri93,Wei90,Bar94,Bar03,Doo92,Lip77,Ade82,Hel85,Bri98,Gre97,Mih97,Mic99,Bar99,Ric02,God85,Bar92,Swa92,Glo96,Fuj96,Glo98,Liu99,Sta03}.
The results of molecular states obtained from different models can
naturally be different.  The present model predicts many molecular
states in the combinations of $\{D, D^*, B, B^*\}$ with $\{\bar D,
\bar D^*, \bar B, \bar B^*\}$.  While it is encouraging that the
present phenomenological model gives a molecular $D^+\bar D^{*-}$
state at about the right energy as the 3872 state, further careful
comparisons of different predictions with experiment will be needed to
determine whether the 3782 state is a molecular state.  Close and Page
\cite{Clo03} and Voloshin \cite{Vol03} suggest that if the 3872 state
is a molecular state, its $D$ and $\bar D^*$ components should decay
with a width equal to those of the isolated mesons, and the 3872 state
should also be seen in the invariant mass of of the combination of the
decay products of $D$ and $\bar D^*$. Furthermore, while the present
model gives $D\bar D$ and $B\bar B$ molecular states, the
pion-exchange model \cite{Tor92,Tor03} does not predict $D\bar D$ and
$B\bar B$ molecular states.  A search for $D\bar D$ and $B\bar B$
molecular states will be of interest to distinguish the different
mechanisms that are present in generating the molecular states.

\begin{acknowledgments}
The authors would like to thank Profs.\ T. Barnes, K. F. Liu, and
E. S. Swanson for helpful discussions.  This research was supported in
part by the Division of Nuclear Physics, U.S. Department of Energy,
under Contract No. DE-AC05-00OR22725, managed by UT-Battelle, LLC.
\end{acknowledgments}


\begin{thebibliography}{99}

\bibitem{Bel03} S.-K. Choi $et~al$, the Belle Collaboration,
Phys. Rev. Lett. {\bf 91}, 262001 (2003); K. Abe $et~al.$, the Belle
Collaboration, hep-ex/0308029; S.-K. Choi, $et~al$, the Belle
Collaboration, hep-ex/0309032.

\bibitem{Bau03} Talk presented by G. Bauer at the Second Heavy
Quarkonium Workshop, Fermilab, September 20-22, 2003; D. Acosta
$et~al$, the CDFII Collaboration, Fermilab-Pub-03/393-E,
hep-ex/0312021.


\bibitem{Tor03}
N. A. T\" ornqvist, 
hep-ph/0308277.

\bibitem{Clo03}
F. Close and P. R. Page,
hep-ph/0309253.

\bibitem{Pak03}
S. Pakavasa and M. Suzuki,
Phys. Lett. {\bf B579},  74 (2004),
hep-ph/0309294.

\bibitem{Vol03}
M. B. Voloshin, hep-ph/0309307

\bibitem{Bib03} 
P. Bicudo, G. M. Marques, hep-ph/0310008

\bibitem{Eic03} E. Eichten, talk presented at the Second Heavy
Quarkonium Workshop, Fermilab, September 20-22, 2003.

\bibitem{Cha03} K. T. Chao, talk presented at the Second Heavy
Quarkonium Workshop, Fermilab, September 20-22, 2003.

\bibitem{Bra03}
E. Braaten and M. Kusunoki, hep-ph/0311147.

\bibitem{Swa03} E. S. Swanson, Physics Letters (in press),
hep-ph/0311229.

\bibitem{Bar03a} T. Barnes and S. Godfrey, Phys. Rev. D {\bf 69},
054008, hep-ph/0311162.

\bibitem{Yua03}
C. Z. Yuan, X. H. Mo, P. Wang,
Phys. Lett. {\bf B579},  74 (2004).

\bibitem{Eic80}
 E. Eichten, K. Gottfried, T. Kinoshita, K. D. Lane and T. M. Yan,
 Phys. Rev. {\bf D21}, 203 (1980).

\bibitem{Buc81}
W. Buchm\" uller and S-H.H. Tye, Phys. Rev. {\bf D24}, 132 (1981).

\bibitem{Eic02}
E. J. Eichten, K. Lane, and C. Quigg, Phys. Rev. Lett. {\bf 89}, 162002
(2002).

\bibitem{Ruj77}
A. De Rujula, H. Georgi and S. L. Glashow, Phys. Rev. Lett. {\bf 38}, 317
(1977); M.B. Voloshin and L.B. Okun, JETP Lett. {\bf 23}, 333 (1976);
M. Bander, G. Shaw, P. Thomas, and S. Meshkov,
Phys. Rev. Lett. {\bf 36}, 695 (1976).

\bibitem{Jaf76} R. J. Jaffe, Phys. Rev. {\bf D15}, 267 (1976), $ibid$
{\bf D15}, 281 (1976).

\bibitem{Dol74} 
A. D. Dolgov, L. B. Okun, and V. I. Zakharov,
Phys. Lett. {\bf 49B}, 455 (1974).

\bibitem{Liu80}
K. F. Liu and C. W. Wong, Phys. Rev. {\bf D21}, 1350 (1980).

\bibitem{Won80}
C. W. Wong and K. F. Liu, Phys. Rev. {\bf D21}, 2039 (1980).

\bibitem{Wei82}
J. Weinstein and N. Isgur, Phys. Rev. Lett. {\bf 48}, 659 (1982);
J. Weinstein and N. Isgur, Phys. Rev. {\bf D27}, 588 (1983);
J. Weinstein and N. Isgur, Phys. Rev. {\bf D41}, 2236 (1990).

\bibitem{Tor92}
N. A. T\" ornqvist, Phys. Rev. Lett. {\bf 67}, 556 (1992); 
N. A. T\" ornqvist, Z. Phys. {\bf C61}, 525 (1994).

\bibitem{Man92}
A. V. Manohar and M. B. Wise, Nucl. Phys. {\bf B 399}, 17 (1993).

\bibitem{Eri93}
T. E. O. Ericson and G. Karl, Phys. Lett. {\bf B309}, 426 (1993).

\bibitem{Wei90}
J. Weinstein and N. Isgur, Phys. Rev. {\bf D41}, 2236 (1990).

\bibitem{Bar94}
T. Barnes, Talk presented at the XXIX Recontres de Moriond, Meribel,
France, 19-26 March 1994 hep-ph/9406215.

\bibitem{Bar03}
T. Barnes, F. E. Close, and H. J. Lipkin,
Phys. Rev. {\bf D68}, 054006,  2003.

\bibitem{Doo92}
K. Dooley, E. S. Swanson, T. Barnes, Phys. Lett. {\bf B275}, 478 (1992).

\bibitem{Lip77} H. J. Lipkin, Phys. Lett. {\bf 70B}, 113 (1977);
H. J. Lipkin, Phys. Lett. {\bf 172B}, 242 (1986); N. Isgur and
H. J. Lipkin, Phys. Lett. {\bf 99B}, 151 (1981).


\bibitem{Ade82}
J. P. Ader, J.M. Richard, and P. Taxil,  Phys. Rev. {\bf D25},  2370 (1982).

\bibitem{Hel85}
L. Heller and J.A. Tjon, Phys. Rev. {\bf D32},  755 (1985).

\bibitem{Bri98}
D. M. Brink and F. Stancu, Phys. Rev. {\bf D57},  6778 (1998).

\bibitem{Gre97} A. M. Green and P. Pennanen, Phys. Lett. {\bf B426},
243 (1998); A. M. Green and P. Pennanen, Phys. Rev. {\bf C57}, 3384
(1998).

\bibitem{Mih97}
A. Mihaly et al., Phys. Rev. {\bf D55},  3077 (1997).

\bibitem{Mic99}
C. Michael and P. Pennanen, 
Phys. Rev. {\bf D60}, 054012 (1999).

\bibitem{Bar99}
T. Barnes, N. Black, D. J. Dean, and E. S. Swanson,
Phys. Rev. {\bf C60}, 045202 (1999).

\bibitem{Ric02}
J.-M. Richard, hep-ph/0212224.

\bibitem{God85}
S. Godfrey and N. Isgur
Phys. Rev. {\bf D32}, 189-231 (1985) 

\bibitem{Bar92}
T. Barnes and E. S. Swanson, Phys. Rev. {\bf D46}, 131 (1992).

\bibitem{Swa92}
E. S. Swanson, Ann. Phys. (N.Y.) {\bf 220}, 73 (1992).

\bibitem{Won02}
C. Y. Wong, E. S. Swanson, and T. Barnes, Phys. Rev. {\bf C65}, 014903 (2002).

\bibitem{Bar03b}
T. Barnes, E. S. Swanson, C. Y. Wong, and X.-M. Xu,
Phys. Rev. {\bf C68},  014903 (2003).

\bibitem{Won00} 
C. Y. Wong, E. S. Swanson, and T. Barnes,
Phys. Rev. {\bf C C62}, 045201 (2000).

\bibitem{Glo96}
L. Ya. Glozman and D. O. Riska, Phys. Rep. {\bf 268}, 263 (1996).

\bibitem{Fuj96}
Y. Fujiwara, C. Nakamoto, and Y. Suzuki,
Phys. Rev. {\bf C54}, 2180 (1996).

\bibitem{Glo98}
L. Ya. Glozman, W. Plessas, K. Varga, and R. F. Wagenbrunn,
Phys. Rev. {\bf D58}, 094030 (1998).

\bibitem{Liu99}
K. F. Liu $et~al.$, Phys. Rev. {\bf D59}, 112001 (1999).

\bibitem{Sta03}
Fl. Stancu and D. O. Riska, 
hep-ph/0307010.

\bibitem{Lip82}
H. J. Lipkin, {\bf 113B}, 490 (1982).

\bibitem{Gre81}
O. W. Greenberg and H. J. Lipkin,
Nucl. Phys. {\bf A370}, 349 (1981).

\bibitem{App77} T. Appelquist, M. Dine, and I. J. Muzinich,
 Phys. Rev. {\bf D17}, 2074 (1977). 

\bibitem{Pes79}
M. E. Peskin, Nucl. Phys. {\bf B156}, 365 (1979).

\bibitem{Bha79a}
G. Bhanot and M. E. Peskin, Nucl. Phys. {\bf B156}, 391 (1979).

\bibitem{Bha79b}
G. Bhanot, W. Fischler, and S. Rudaz, Nucl. Phys. {\bf B155}, 208 (1979).


\bibitem{Alm60}
E. Almqvist, D. A. Bromley, and J. A. Kuehner, Phys. Rev. Lett. {\bf 4}, 515
(1960).

\bibitem{Kon79}
Y. Kondo, Y. Abe, and T. Matsuse, Phys. Rev. {\bf C19}, 1356 (1979). 


\bibitem{Sat84}
G. R. Satchler, {\sl Direct Nuclear Reactions}, (Oxford University
Press, Oxford, 1983);  G. R. Satchler and W. G. Love, Phys. Rep. {\bf 55},
183 (1979).

\bibitem{Won01} C. Y. Wong and H. W. Crater, Phys. Rev. {\bf C63},
044907 (2001).

\bibitem{others}
H. Lipkin,  Phys. Lett.  {\bf B45}, 267 (1974);
R. S. Willey, Phys. Rev. {\bf D18}, 270 (1978); 
P. M. Fishbane and M. T. Grisaru, Phys. Lett. {\bf B74}, 98 (1978); 
G. Feinberg and J. Sucher, Phys. Rev. {\bf D20}, 1717 (1979); 
S. Matsuyama and H. Miyazawa, Prog. Theo. Phys. {\bf 61}, 942 (1979); 
Y. Fujii and K. Mima, Phys. Lett. {\bf B79}, 138 (1978).

\bibitem{Cas47}
H. B. G. Casimir and D. Polder, Phys. Rev. {\bf 73}, 360 (1947).

\bibitem{Luc91}
J. Lucinda, J. Phys. {\bf  A 24}, 1759 (1991).

\bibitem{Bal00}
G. S. Bali, Phys. Rev.  {\bf D62}, 114503 (2000)

\bibitem{Zha03} 
W. N. Zhang and C. Y. Wong, Phys. Rev. {\bf C68}, 035211 (2003).

\bibitem{Won99} C. Y. Wong, Phys. Rev. {\bf D60}, 114025 (1999).

\bibitem{Won94}
C. Y. Wong, {\it Introduction to High-Energy Heavy-Ion Collisions},
World Scientific Publishing Company, 1994.

\bibitem{Sch51}
J. Schwinger, Phys. Rev. {\bf 82}, 664 (1951).

\bibitem{Pet75}
F. Petrovich, Nucl. Phys. {\bf A251}, 143 (1975).


\end{thebibliography}
\end{document}